# Directly visualizing the sign change of *d*-wave superconducting gap in $Bi_2Sr_2CaCu_2O_{8+\delta}$ by phase-referenced quasiparticle interference

Qiangqiang Gu[1], Siyuan Wan[1], Qingkun Tang[1], Zengyi Du, Huan Yang[1], Qiang-Hua Wang[1], Ruidan Zhong[2], Jinsheng Wen[2], G. D. Gu[2] & Hai-Hu Wen[1]

[1]National Laboratory of Solid State Microstructures and Department of Physics, Center for Superconducting Physics and Materials, Collaborative Innovation Center for Advanced Microstructures, Nanjing University, Nanjing 210093, China.

[2]Condensed Matter Physics and Materials Science Department, Brookhaven National Laboratory, Upton, New York 11973, USA.

These authors contributed equally: Qiangqiang Gu, Siyuan Wan, and Qingkun Tang. Correspondence and requests for materials should be addressed to H.Y. (email: huanyang@nju.edu.cn) or to H.-H.W. (email: hhwen@nju.edu.cn).

**The superconducting state is achieved by the condensation of Cooper pairs and is protected by the superconducting gap. The pairing interaction between the two electrons of a Cooper pair determines the superconducting gap function. Thus, it is very pivotal to detect the gap structure for understanding the mechanism of superconductivity. In cuprate superconductors, it has been well established that the superconducting gap may have a *d*-wave function $\Delta = \Delta_0\cos2\theta$. This gap function has an alternative sign change by every $\pi/2$ in the momentum space when the in-plane**

**azimuthal angle $\theta$ is scanned. It is very hard to visualize this sign change. Early experiments for recommending or proving this *d*-wave gap function were accomplished by the specially designed phase sensitive measurements based on the Josephson effect. Here we report the measurements of scanning tunneling spectroscopy in one of the model cuprate system $Bi_2Sr_2CaCu_2O_{8+\delta}$ and conduct the analysis of phase-referenced quasiparticle interference (QPI). Due to the unique quasiparticle excitations in the superconducting state of cuprate, we have seen the seven basic scattering vectors that connect each pair of the terminals of the banana-shaped contour of constant quasiparticle energy (CCE). The phase-referenced QPI clearly visualizes the sign change of the *d*-wave gap. Our results illustrate a very effective way for determining the sign change of unconventional superconductors.**

## Introduction

In superconductors, the charge carriers are Cooper pairs which carry the elementary charge of 2*e*. This can be easily concluded from the measurements of the quantized flux $\Phi_0 = h/2e = 2.07 \times 10^{-15}$ Wb based on the Ginzburg-Landau theory. The core issue and also the most hard problem for understanding the mechanism of superconductivity is how the two conduction electrons are bound each other to form a pair, the so-called Cooper pair. In the theory of Bardeen-Cooper-Schrieffer (BCS), it was predicted that the Cooper pairs in some superconductors, mainly in elementary metals and alloys, are formed by exchanging the virtue vibrations of the atomic lattice, namely phonons. The two electrons in the original paired state (**k**, −**k**) will be scattered to another paired state (**k**′, −**k**′). This scattering will lead to the attractive interaction $V_{k,k'}$, and the electron bound state will be formed with the help of suitable Coulomb screening. Condensation of these electron pairs will lead to superconductivity. This condensate is protected by an energy gap $\Delta(k)$ which prevents the breaking of Cooper pairs. Usually the gap is a momentum dependent function which is closely related to the pairing interaction function $V_{k,k'}$. For example, in above mentioned pair scattering picture, one can easily derive the function

$$\Delta(k) = -\sum_{k'} V_{k,k'} \frac{\Delta(k')}{2E(k')} \tanh\frac{E(k')}{2k_{\mathrm{B}}T}. \quad (1)$$

Here $E(k) = \sqrt{\varepsilon^2 + \Delta^2(k)}$ with $\varepsilon$ the kinetic energy of the quasiparticles counting from the Fermi energy $E_F$. If this pairing process can be lend to other unconventional superconductors, the sign of the gap $\Delta(k)$ would change if the pairing function $V_{k,k'}$ is positive.

For cuprate superconductors, it has been well documented that the gap has a *d*-wave form $\Delta=\Delta_0\cos 2\theta$. This basic form of the gap was first observed by experiments of angle resolved photo-emission spectroscopy (ARPES) without sign signature[1,2], and later supported by many other experiments, such as thermal conductivity[3], specific heat[4,5], scanning tunneling microscopy (STM)[6,7], neutron scattering[8,9] and Raman scattering[10], etc. Although some of the techniques mentioned above may involve the sign change of the gap, such as the inelastic neutron scattering and STM measurements, but they cannot tell how the gap sign changes explicitly across the momentum space.

In the model system of cuprate superconductor $Bi_2Sr_2CaCu_2O_{8+\delta}$(Bi-2212), the low energy excitations have been measured very carefully by STM yielding the seven scattering vectors or spots in the Fourier transformed (FT-) quasi-particle interference (QPI) pattern[11]. These seven scattering vectors were explained very well with the Fermi arc picture as evidenced by ARPES measurements[12]. Combining the magnetic effect on the QPI intensities of the different vectors, Hanaguri *et al.*[13] showed the well consistency of a *d*-wave gap, if assuming that the magnetic vortices behave like magnetic scattering centers and that leads to the enhancement (suppression) of the joint QPI intensity of two momenta ($\mathbf{k}_1, \mathbf{k}_2$) with the gaps of the same (opposite) signs. The sign problem of a *d*-wave gap was also resolved by the experimental devices based on the Josephson effect[14-16]. Therefore *d*-wave superconducting gap has been well concluded in cuprate superconductors by using many experimental tools. In this paper we report the experiments and analysis based on the impurity induced bound state (IBS-) QPI method[17,18] on Bi-2212. Our results illustrate the direct visualization of the sign change of the *d*-wave gap in bulk Bi-2212.

## Results

**Topography and tunneling spectra measured on Bi-2212**. Figure 1a shows a typical topographic image of the BiO surface of Bi-2212 with an atomic resolution after cleaverage. The supermodulations in the real space can be clearly witnessed on the surface with a period of about 2.54 nm. We can also recognize a square lattice structure with the lattice constant $a_0$ about 3.5 Å, even with the appearance of the supermodulations. In Fig. 1b, we present a sequence of tunneling spectra measured along the arrowed line in Fig. 1a. The spectra show the ‘V’-shaped bottoms near zero bias, which reveals the intrinsic feature of the gap nodes in optimally doped Bi-2212. One can also clearly find the spatial variation of the coherence-peak positions, being consistent with previous reports[6,19]. We then compute the averaged spectrum of the 30 spectra shown in Fig. 1b, and plot it in Fig. 1c. The finite zero-bias differential conductance implies the effect of the impurity scattering[6,20] in the case of a nodal gap. We then use the Dynes model[21] with a *d*-wave gap function $\Delta(\theta) = \Delta_0 \cos 2\theta$ to fit the average spectrum, and the fitting result is also shown in Fig. 1c by the red line. The resultant fitting yields the parameters of gap maximum $\Delta_0$ = 40 meV, and the scattering rate $\Gamma$ = 2.6 meV. One can see that the fitting curve with the *d*-wave gap function captures the major characteristics except for the feature of the coherence peak in negative-bias side. It should be noted that the gap maximum equals approximately to the coherence-peak energy in a *d*-wave superconductor when the scattering rate is small. Following this conclusion, we carried out a statistical analysis of the gap maximum values from the coherence-peak energies of about 200 spectra measured in our experiments on optimally doped Bi-2212 samples, and the related histogram of the gap distribution is presented in Fig. 1d. One can see that the distribution of $\Delta_0$ behaves as a Gaussian shape with an averaged gap maximum $\overline{\Delta}_0$ = 43 meV obtained from the fitting.

**Identification of different scattering wave-vectors in FT-QPI pattern**. Figure 2a shows a topographic image with the dimensions of 56 nm×56 nm. In this area, we measured the QPI images $g(\mathbf{r},E)$ at different energies, and then obtained the FT-QPI pattern $g(\mathbf{q},E)$ in **q**-space through the Fourier transform. Figure 2b shows a typical FT-QPI pattern at 20 meV. Since we have done the lattice correction[22] process, the four Bragg peaks are very sharp near the red crosses marked in Fig. 2b. Based on the octet model[23], the terminals of the banana-shaped CCE has relatively higher density of states (DOS). Therefore, the major scattering will occur among these hot spots. As expected, we can clearly distinguish all the primary scattering wave vectors $\mathbf{q}_i$ ($i$ = 1 to 7) indicated by dashed circles in the FT-QPI pattern. Besides, two additional pairs of spots, which are indicated by solid-line circles, are originated from the supermodulations[23]. It should be noted that the $\mathbf{q}_7$ spots at the horizontal direction are not influenced by the supermodulations. Figure 2c shows a schematic plot of the contours at some energy below the superconducting gap maximum in a typical cuprate superconductor like Bi-2212, and the DOS at the terminals are set to be maximum. By assuming a certain width for the arcs in Fig. 2c, we then apply the self-correlation and show the simulated FT-QPI result in Fig. 2d. One can see that the simulated patterns are similar to those in the experimental FT-QPI pattern, which can help identify different primary scattering wave vectors. In a *d*-wave superconductor, the superconducting gap changes its sign in **k**-space along the Fermi surfaces, and we use different background colours to show this sign-change feature in Fig. 2c. With the aid of light pink and light blue colours for different gap signs, we can divide the scattering wave vectors into two types, i.e., gap-sign-reversed scatterings (with scattering wave vectors of $\mathbf{q}_2$, $\mathbf{q}_3$, $\mathbf{q}_6$, and $\mathbf{q}_7$) and gap-sign-preserved ones ($\mathbf{q}_1$, $\mathbf{q}_4$, and $\mathbf{q}_5$).

**Energy evolution for different scattering vectors**. It is known that the Bogoliubov quasiparticles will interfere with each other in the presence of impurities in a superconductor, giving rise to the Friedel-like oscillations of local DOS (LDOS). Figure 3a-f present the QPI images measured at different energies. We can see clear standing waves in these QPI images, which are due to the scattering by some impurities or defects in the sample such as randomly distributed oxygen deficiencies[20]. The longitudinal periodic modulations as shown in the topography of Fig. 2a mainly contribute the spots enclosed by the solid-line circles along $q_y$ in Fig. 2b. While the other spots in Fig.2b should be contributed by the scatterings between the terminals of CCE. We then do the Fourier transform to QPI images and show three sets of them in Fig. 3g-i, and the measured QPI images and corresponding FT-QPI patterns at other energies are shown in Supplementary Fig. 1. One can clearly recognize the scattering spots from supermodulations and seven characteristic scatterings in the FT-QPI patterns. It should be noted that the coordinates of supermodulation spots in FT-QPI patterns are non-dispersive in energy, which can also be taken as a judgement for the origin of the supermodulation. However, the **q**-vectors of seven characteristic scattering spots seem to be dispersive, which can be naturally illustrated by the octet model[23].

To obtain the information of superconducting gaps, we then analyze the energy dispersions of the characteristic scattering wave vectors. The intensities of the spots corresponding to $\mathbf{q}_1$ and $\mathbf{q}_7$ scattering vectors are stronger than the spots corresponding to other scattering vectors, and the central coordinates of these spots can be obtained by the fitting to a two-dimensional Lorentzian function[11]. Fortunately, these two scattering wave vectors can provide enough information to fix the coordinates of the octet terminals at various energies within the superconducting gap maximum. The resultant positions of the octet terminals are shown in the inset of Fig. 4a, and the data allow us to construct part of the Fermi surface. The solid curve in the inset is a fitting result to the positions of the octet terminals by a circular arc. The curve, which is intentionally cut off by the dashed line[24], represents the contour of the Fermi surface of Bi-2212 at this doping level. From the inset of Fig. 4a, we can also define the angle of these octet terminals to the (0, π) to (π, π) direction. With the combination of the defined angle $\theta$ and the measured energies as the absolute values of superconducting gaps for the octet ends, we can obtain the angular dependent superconducting gap $\Delta(\theta)$ along the Fermi surface and show it in Fig. 4a. The obtained $\Delta(\theta)$ data are fitted by two different *d*-wave gap functions, i.e., $\Delta_1(\theta)$ = 39cos2$\theta$ meV and $\Delta_2(\theta)$ = 45.6 (0.9cos2$\theta$+0.1cos6$\theta$) meV, and the latter shows a better consistence with the experimental data. Therefore, the gap function in optimally doped Bi-2212 may have a higher order in addition to the *d*-wave component, which is consistent with previous report[23]. Looking back to other scattering spots, the intensity of $\mathbf{q}_4$ scattering spot is very weak. The positions of $\mathbf{q}_2$ and $\mathbf{q}_6$ scatterings are equivalent with only the exchange of $k_x$ and $k_y$ coordinates. Thus, we focus only on five sets of visible characteristic scattering spots with the center coordinates obtained by the fittings to two-dimensional Lorentzian functions. The obtained energy dispersions of these characteristic wave vectors are presented in Fig. 4b, being consistent with previous reports[7,19,23]. The solid curves in Fig. 4b represent the predicted energy dispersions based on the Fermi surface in the inset of Fig. 4a and the gap function $\Delta_2(\theta)$. One can see that the calculated curves agree well with the experimental data, which indicates the validity of the octet model and the *d*-wave superconducting gap.

**Theoretical approach of using the IBS-QPI method to a *d*-wave superconductor**. The PR-QPI method was first theoretically proposed by Hirschfeld, Altenfeld, Eremin, and Mazin in a two-gap superconductor[25], which was successfully applied to prove the gap-sign-change in iron-based superconductors[26,27]. The recently proposed IBS-QPI method[17,18] is designed for judging the gap-sign problem in iron-based superconductor LiFeAs, and is also successfully used to confirm the gap sign-change in $(Li_{1-x}Fe_x)OHFe_{1-y}Zn_ySe$ from our previous work[28]. In this method, the QPI image $g(\mathbf{r},E)$ is measured in an area with a non-magnetic impurity sitting at the center of the field of view (FOV). Then the FT-QPI data, which comes from the Fourier transform of $g(\mathbf{r},E)$, are complex parameters containing the phase information, namely $g(\mathbf{q},E) = |g(\mathbf{q},E)|e^{i\varphi_g(\mathbf{q},E)}$. Then the phase-referenced (PR-) QPI signal can be extracted from the phase difference between positive and negative bound state energies, that is defined by

$g_r(\mathbf{q},+E) = |g(\mathbf{q},+E)|,$ (2)

$g_r(\mathbf{q}, -E) = |g(\mathbf{q}, -E)|\cos(\varphi_{\mathbf{q},-E} - \varphi_{\mathbf{q},+E}).$ (3)

Here $g_r(\mathbf{q},+E)$ should be always positive from the definition, and $g_r(\mathbf{q}, -E)$ should be negative near the bound state energy for the scattering involving the sign-reversal gaps at $\mathbf{k}_1$ and $\mathbf{k}_2$ ($\mathbf{q} = \mathbf{k}_1$-$\mathbf{k}_2$). That conclusion was drawn by the simulation for a nodeless superconductor when the gap changes its sign for different Fermi pockets in LiFeAs[17,18]. Our studies in $(Li_{1-x}Fe_x)OHFe_{1-y}Zn_ySe$ reveal that this method can also work for concentric two circle like Fermi surfaces when the gaps on them have opposite signs[28].

In cuprates, the gap value varies continuously and gap nodes appear in the nodal direction ($\Gamma$-Y) on the Fermi surface. The FT-QPI patterns have already been calculated in several previous works on cuprate systems[29-35]. However, it is very curious to check whether this phase-referenced IBS-QPI method is still applicable theoretically in a *d*-wave superconductor. Before checking, it seems that there are no bound state peaks on the tunneling spectrum, but the bound state peaks seem to be necessary for the IBS-QPI technique. In cuprates, the intrinsic nanoscale electronic disorder such as oxygen vacancies or crystal defects can act as the scattering centers, which will influence the tunneling spectra[20]. According to the previous calculation[36], if the scattering potential is small, the bound state peaks may merge to the coherence peaks. Obviously, the coherence-peak positions measured in Bi-2212 has a spatial distribution, as illustrated in Fig. 1 and from previous reports[6,19]. The possible reason for this is the mixture of different impurity bound states. To prove this issue, we do the theoretical calculations by a standard T-matrix method[29] with the details described in Method part. Supplementary Fig. 2a shows the calculated angle dependent superconducting gap and Fermi surface, which is consistent with the experimental data. The tunneling spectra with or without non-magnetic impurity is shown in Supplementary Fig. 2b. One can see that the non-magnetic impurity induced bound state peak merge to the coherence-peaks when the scattering potential is as small as 20 meV. Combined with theoretical calculations, it is reasonable to believe that there may be many impurities on the surface with very small scattering potential, and the bound state peaks are mixed with the coherence-peaks. The simulation results following the IBS-QPI method for a single impurity are shown in Supplementary Fig. 3a-e. One can see that the PR-QPI signal for the gap-sign-preserved scatterings ($\mathbf{q}_1$, $\mathbf{q}_4$, and $\mathbf{q}_5$) are positive, while the signal for the gap-sign-reversed ones ($\mathbf{q}_2$, $\mathbf{q}_3$, $\mathbf{q}_6$, and $\mathbf{q}_7$) are negative, which gives a sharp contrast. The simulated result in a *d*-wave superconductor seems to be consistent with the situation in the $s^{\pm}$ superconductor LiFeAs (Ref. 17,18).

**Multi-IBS-QPI method applied on experimental data in Bi-2212.** Figure 5a-f shows the PR-QPI patterns at different negative energies. The PR-QPI patterns at positive energies are not shown here, because all the signals should be positive without any extra phase information according to Eq. 2. At energies from −6 to −12 meV, the $g_r(\mathbf{q},-E)$ signals of $\mathbf{q}_1$ and $\mathbf{q}_7$ are clear and easily recognized. When the energy is lowered to below −15 meV, all the characteristic scattering spots become clear and can be easily recognized. We then calculate the average value to the signals in the areas within the dashed-circles or ellipses in Fig. 5d-f, and the histograms of the average intensities per pixel corresponding to different scattering channels are shown in Fig. 5g-i. The intensities corresponding to $\mathbf{q}_1$, $\mathbf{q}_4$ and $\mathbf{q}_5$ spots are positive, while those corresponding to $\mathbf{q}_2$, $\mathbf{q}_3$, $\mathbf{q}_6$ and $\mathbf{q}_7$ spots are negative. As a result, we argue that those **k** points in the momentum space connected by $\mathbf{q}_1$, $\mathbf{q}_4$ or $\mathbf{q}_5$ have a sign-preserved superconducting order parameters, and those connected by $\mathbf{q}_2$, $\mathbf{q}_3$, $\mathbf{q}_6$ or $\mathbf{q}_7$ have a sign-reversed ones. This conclusion is consistent very well with our theoretical calculation above by using a *d*-wave gap function. Since the theoretical model used for the Fermi surface is rather rough, it is thus reasonable to see different intensities of the corresponding spots of the PR-QPI signal between the experimental data (Fig. 5g-i) and the calculated result (Supplementary Fig. 3j).

**Discussion**

Although consistency has been found between the experimental data and theoretical calculations by using the IBS-QPI technique, however, one may argue that the original phase-referenced QPI method[17,18] is specially designed for the case of a single impurity. In the case of Bi-2212, many randomly distributed weak impurities prevent us from finding an area with a well-isolated impurity, and the LDOS modulation induced by an impurity could be interfered by other nearby impurities. As a result, we can only measure in a large area with multiple impurities, giving rise to rambling standing waves in the conductance mappings as shown in Fig. 3a-f. Our primary concern is whether the phase message of the order parameter can be effectively extracted in the case of multiple impurities.

The initial theoretical work of the IBS-QPI technique[17] has actually provided a treatment for the multi-impurity system by applying the phase correction[37] to the FT-QPI pattern $g(\mathbf{q},E)$ as

$$g_{\mathrm{m}}(\mathbf{q},E)=\frac{g(\mathbf{q},E)}{C(\mathbf{q})}=\frac{|g(\mathbf{q},E)|}{|C(\mathbf{q})|}e^{i[\varphi_g(\mathbf{q},E)-\varphi_C(\mathbf{q})]}. \quad (4)$$

Here the correction term $C(\mathbf{q})=\sum_j e^{-i\mathbf{q}\cdot\mathbf{R}_j}$ with $\mathbf{R}_j$ the location of the $j^{\mathrm{th}}$ impurity. The PR-QPI signals can be obtained by applying Eqs. 2 and 3 to the corrected FT-QPI $g_{\mathrm{m}}(\mathbf{q},E)$ data obtained by Eq. 4 (Ref. 17). However, this correction seems to be difficult to apply in Bi-2212 because of the uncertainness of the impurity positions. From the QPI images shown in Fig. 3, one can clearly see that there should be many weak impurities on the surface and it is very difficult to determine the exact coordinates of these impurities. Therefore, the phase-correction is impossible to apply in Bi-2212. However, considering the correction factor $C(\mathbf{q})=\sum_j e^{-i\mathbf{q}\cdot\mathbf{R}_j}$ is energy independent, it will affect the phase of $g_{\mathrm{r}}(\mathbf{q},+E)$ and $g_{\mathrm{r}}(\mathbf{q},-E)$ simultaneously. The new phase after correction can be written as $\varphi_{\mathrm{m}}(\mathbf{q},E)=\varphi_g(\mathbf{q},E)-\varphi_C(\mathbf{q})$, and the phase difference $\varphi_{\mathrm{m}}(\mathbf{q},-E)-\varphi_{\mathrm{m}}(\mathbf{q},+E)=\varphi_{\mathrm{g}}(\mathbf{q},-E)-\varphi_{\mathrm{g}}(\mathbf{q},+E)$ remains unchanged in the multi-impurity system. It means that the sign of PR-QPI signal at the negative energy for multiple impurities will be exactly the same to the one for a single impurity. We can get support also from the theoretical calculations. To illustrate that, we assume a case for 60 impurities randomly distributed on the surface, and the resultant LDOS image after calculation is shown in Supplementary Fig. 3f. The PR-QPI signals for all the scattering wave-vectors (Supplementary Fig. 3j) are of the same sign but different values compared with the situation of the single impurity (Supplementary Fig. 3e). We have repeated the simulations for 100 times with different distributions of the impurities, and can obtain the same conclusion. Hence, this phase-sensitive method for multi-impurities with bound states near the coherence-peaks can provide us a feasible and easy way to detect the sign problem on the unconventional superconductors.

As presented above, we use the multi-IBS-QPI method to prove the superconducting gap reversal in optimally doped Bi-2212, which is consistent very well with the *d*-wave gap structure. However, with these results, one may argue that perhaps the sign-preserved gap may also give rise to such changes of the phase difference between the positive and negative energies. It has been calculated and argued that, the scattering from a non-magnetic impurity can barely induce a impurity bound state locating near the coherence-peak in an isotropic-*s*-wave superconductor[38]. If the gap is highly anisotropic, or say its value varies in a wide range, there may be some impurity induced bound states even if the superconducting gap is nodeless. We do further simulations for the superconductors with different gap functions, for examples, a nodal but sign-preserved gap $\Delta(\mathbf{k}) = 23|\cos k_x - \cos k_y|$ (meV) and a nodeless gap $\Delta(\mathbf{k}) = 23\left|\cos k_x - \cos k_y\right| + 2$ (meV). The same scattering scalar potential $V_\mathrm{s} = 20$ meV is used for the non-magnetic impurity. The resultant tunneling spectra are shown in Supplementary Figs. 2c and 2d, respectively. One can see that the non-magnetic impurities only slightly shift the position of the coherence-peaks, and have negligible influence on the LDOS near zero bias. We also calculate the PR-QPI images for the single and multi- impurity situation with different forms of the sign-preserved superconducting gaps mentioned above, and the related simulations are shown in Supplementary Figs. 4 and 5. One can find that the results for the multi-impurity situation are always similar to the case for a single impurity. Furthermore, all the PR-QPI signals for seven characteristic scattering spots are positive in this case. Therefore, we exclude the possibility of the sign-preserved gap in Bi-2212.

Another argument may be that the impurities could be magnetic ones instead of non-magnetic ones, and then the magnetic potential should be considered in the scattering[38]. This is actually no relevant for the optimally doped Bi-2212 since, as far as we know, no magnetic impurities with even moderate scattering potentials have been reported in literatures. The calculated PR-QPI signal was shown to be always positive for magnetic impurities in $s^{++}$ situation as proposed for LiFeAs[17,18]. We have also done the calculation here for the superconductors with a sign-preserved gap with magnetic impurities, and find that the PR-QPI signals for seven characteristic scattering spots are of the same signs.

The successful application of the multi-IBS-QPI method to a *d*-wave superconductor may be attributed to the wide range of superconducting gap values. In this situation, although the scattering wave vectors have obvious energy dispersions, there should be some hotspots of the Bogoliubov quasiparticles at any energy within the superconducting gap maximum. With the enhancement effect of joint DOS of Bogoliubov particles near the coherence peaks, we can easily obtain the PR-QPI signal with the assistance of widely distributed impurities of weak scattering potentials.

Although the peaks from the impurity induced bound states merge to the coherence peaks, the effect of the bound states extends to all the energy of the spectra from the simulated tunnelling spectra in Supplementary Fig. 2b. Indeed, the scattering potentials are different for different impurities, but the scatterings at different energies are always there. In this framework, we clearly prove the sign-changing *d*-wave pairing symmetry in optimally doped Bi-2212. Our experiments and analyzing technique also suggest that this method may also be applicable to any other unconventional superconductors if the gap sign has a reversal. This will provide a more easily accessible method to determine the gap structure of unconventional superconductors.

## Methods

**Sample synthesis and characterization.** Optimally-doped $Bi_2Sr_2CaCu_2O_{8+\delta}$ single crystals were grown by the floating-zone technique[39]. The quality of the sample has been checked by the DC magnetization measurement before the STM measurements. The critical temperature $T_c$ is about 90 K as determined from the DC magnetization measurement.

**STM/STS measurements**. The STM/STS measurements were operated in a scanning tunneling microscope (USM-1300, Unisoku Co., Ltd.) with ultra-high vacuum, low temperature and high magnetic field. The Bi-2212 samples were cleaved at room temperature in an ultra-high vacuum with a base pressure of about $1\times10^{-10}$ torr. The electrochemically etched tungsten tips or the Pt/Ir alloy tips were used for all the STM/STS measurements. A lock-in technique was used for measuring tunneling spectrum with an ac modulation of 1 mV and 987.5 Hz. All the data were taken at 1.5 K.

**Theoretical calculations.** We have employed a single tight-binding band structure similar to the one proposed in the previous report[40], with the energy dispersion given by

$$\varepsilon_k = -2t_1\left(\cos k_x + \cos k_y\right) + 4t_2 \cos k_x \cos k_y - 2t_3(\cos 2k_x + \cos 2k_y) - 2t_4(\cos 2k_x \cos k_y + \cos k_x \cos 2k_y) - \mu. \quad (5)$$

The parameters $(t_1, t_2, t_3, t_4, \mu)$ = (100, 36, 10, 1.5, −155) are used in the calculations with the units of meV. Using a standard T-matrix method[29], we simulate the LDOS around a single impurity and for the case of multiple impurities. In Nambu space, the BCS Hamiltonian is given by

$$H_0 = \begin{pmatrix} \varepsilon_k & \Delta_k \\ \Delta_k^* & -\varepsilon_{-k} \end{pmatrix}. \quad (6)$$

Here we have N impurities located at $\mathbf{r}_1$, $\mathbf{r}_2$, $\mathbf{r}_3$, …, and $\mathbf{r}_N$ ( Set N = 1 reduced to the case of a single impurity). The Green's function in real space[41] can be formulated by

$$G(\mathbf{r}, \mathbf{r}', E) = G_0(\mathbf{r}, \mathbf{r}', E) + \sum_{i,j} G_0(\mathbf{r}, \mathbf{r}_i, E) T(\mathbf{r}_i, \mathbf{r}_j, E) G_0(\mathbf{r}_j, \mathbf{r}', E), \quad (7)$$

where $G_0(\mathbf{r}, \mathbf{r}', E) = \frac{1}{M}\sum_{\boldsymbol{k}} G_0(\boldsymbol{k}, E)\, e^{i\boldsymbol{k}\cdot(\mathbf{r}-\mathbf{r}')}$ with M the numbers of unit cells, $G_0(\boldsymbol{k}, E)$ the unperturbed Green's function in reciprocal space and the many-impurity 2N×2N T-matrix is determined by

$$\mathbf{T}^{-1} = \mathbf{V}^{-1} - \mathbf{G_0}, \tag{8}$$

where

$$\mathbf{G_0} = \begin{pmatrix} G_0(\mathbf{r_1}, \mathbf{r_1}, E) & \cdots & G_0(\mathbf{r_1}, \mathbf{r_N}, E) \\ \vdots & \ddots & \vdots \\ G_0(\mathbf{r_N}, \mathbf{r_1}, E) & \cdots & G_0(\mathbf{r_N}, \mathbf{r_N}, E) \end{pmatrix}, \tag{9}$$

$$\mathbf{V} = \begin{pmatrix} V_1 & 0 & \cdots & 0 \\ 0 & V_2 & \cdots & 0 \\ \vdots\ \vdots & & \ddots & \vdots \\ 0 & 0 & \cdots & V_N \end{pmatrix}. \tag{10}$$

Here, $V_i = V_s\tau_3$, with $V_s$ the scalar potential, $\tau_3$ the third Pauli matrix. Then we can obtain the LDOS given by

$$g(\mathbf{r}, E) = -\frac{1}{\pi}\mathrm{ImTr}[\frac{\tau_0+\tau_3}{2} G(\mathbf{r}, \mathbf{r}, E)], \tag{11}$$

where $\tau_i$ is the Pauli matrix spanning Nambu space. Referring to Eq. 2 and 3, we can get the simulated PR-QPI images as shown in Supplementary Fig. 3 and 4.

**Data availability.** The data that support the plots within this paper and other findings of this study are available from the corresponding author upon reasonable request.

**Acknowledgements**

We acknowledge useful discussions with Tetsuo Hanaguri and Dunghai Lee. We thank Zhen Ke for the programming of lattice correlation. The work was supported by National Key R&D Program of China (grant number: 2016YFA0300401), National Natural Science Foundation of China (NSFC) with the project numbers 11534005, 11604168. Work at Brookhaven was supported by the Office of Basic Energy Sciences (BES), Division of Materials Science and Engineering, U.S. Department of Energy (DOE), through Contract No. de-sc0012704. R. D. Z. and J. S. were supported by the Center for Emergent Superconductivity, an Energy Frontier Research Center funded by BES.

**Author contributions**

The low-temperature STM/STS measurements and analysis were performed by Q.G.,S.Y.W.,Z.Y.D., H.Y. and H.H.W. The samples were grown by G.G., R.D.Z., and J.S.W.. Q.G., H.Y. and H.H.W contributed to the writing of the paper. The theoretical calculation was done by Q.T., Q.G and Q.H.W. H.Y. and H.H.W. are responsible for the final text. H.H.W. coordinated the whole work. All authors have discussed the results and the interpretations.

**Competing financial interests**

The authors declare that they have no competing financial interests.

**Figures and captions**

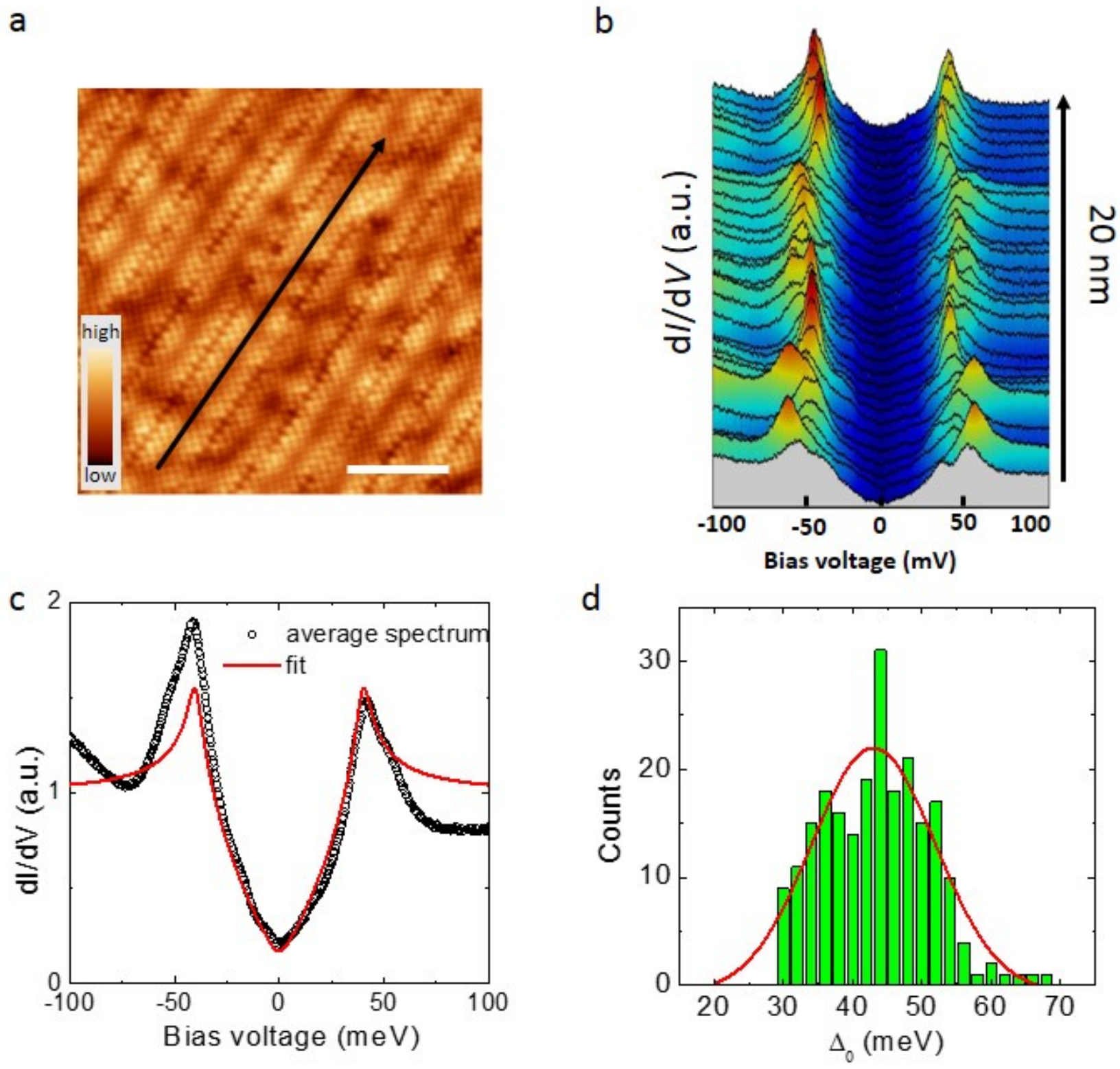

**Fig. 1** Topographic image and tunneling spectra measured on optimally doped Bi-2212. **a** A typical topographic image of the BiO surface after cleavage measured at bias voltage $V_{bias}$ = 50 mV and tunneling current $I_t$ = 100 pA. Scale bar, 5 nm. **b** Tunneling spectra measured along the arrowed line in **a**. **c** The averaged spectrum (open circles) of 30 spectra in **b**. The set-point conditions are $V_{bias}$ = 100 mV and $I_t$ = 100 pA for all the spectra. The solid line shows the Dynes model fitting result with a *d*-wave superconducting gap. **d** The statistics of superconducting gap maxima $\Delta_0$ for all the tunneling spectra measured on Bi-2212 samples, and the gap maximum values are determined from the positions of the coherence-peaks. The red curve is a Gaussian fitting result with the peak position near 43 meV.

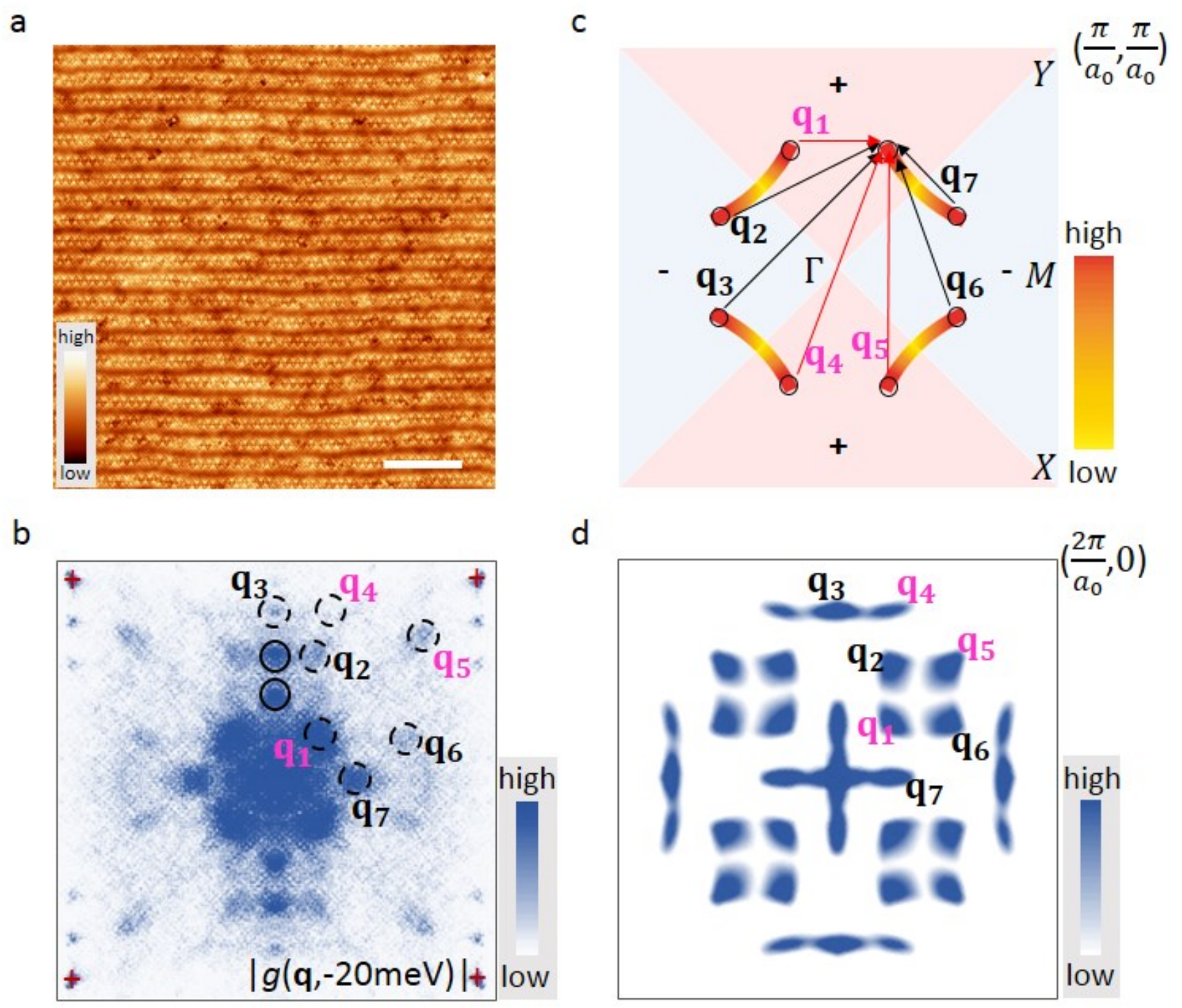

**Fig. 2** Different kinds of scattering wave-vectors in FT-QPI pattern. **a** Topographic image of the BiO plane ($V_{\text{bias}}$ = −200 mV, $I_{\text{t}}$ = 50 pA). Scale bar, 10 nm. **b** FT-QPI pattern ($V_{\text{bias}}$ = −100 mV, $I_{\text{t}}$ = 50 pA) at $E$ = −20 meV based on the QPI image measured in the area of **a**. Four Bragg peak positions are marked by the red crosses. The solid-line circles indicate the scattering spots contributed from the supermodulations, while the dashed circles indicate the primary scattering spots with different **q** vectors. **c** A schematic plot of the contours of constant energy. The DOS along the Fermi surface is set to be **k**-dependent, and the intensities at the octet terminals are the strongest. The different colours of light pink and light blue show the regions with different gap signs for the *d*-wave gap function. The scatterings along black arrows ($\mathbf{q}_2$, $\mathbf{q}_3$, $\mathbf{q}_6$, and $\mathbf{q}_7$) are for sign reversal gaps, while the scatterings along magenta arrows ($\mathbf{q}_1$, $\mathbf{q}_4$, and $\mathbf{q}_5$) are sign-preserved ones. **d** The simulation of FT-QPI by applying the self-correlation to **c**.

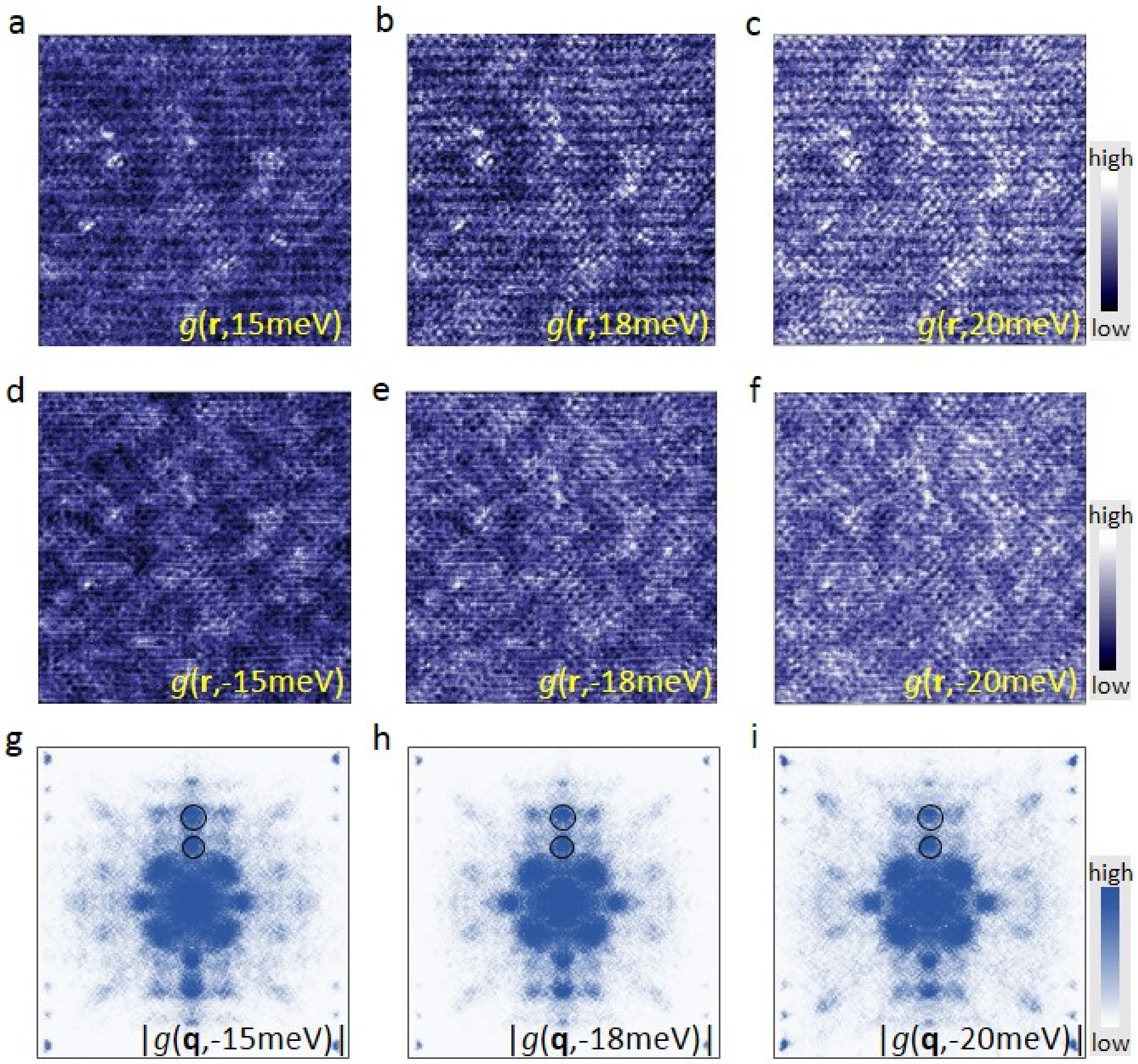

**Fig. 3** QPI and FT-QPI patterns measured on Bi-2212. **a**-**f** Differential conductance mappings at different energies. **g**-**i** Three corresponding FT-QPI patterns of **d**-**f**, respectively. The supermodulations contribute to the FT-QPI intensity surrounded by the solid-line circles, and these spots seem to have no energy dispersion.

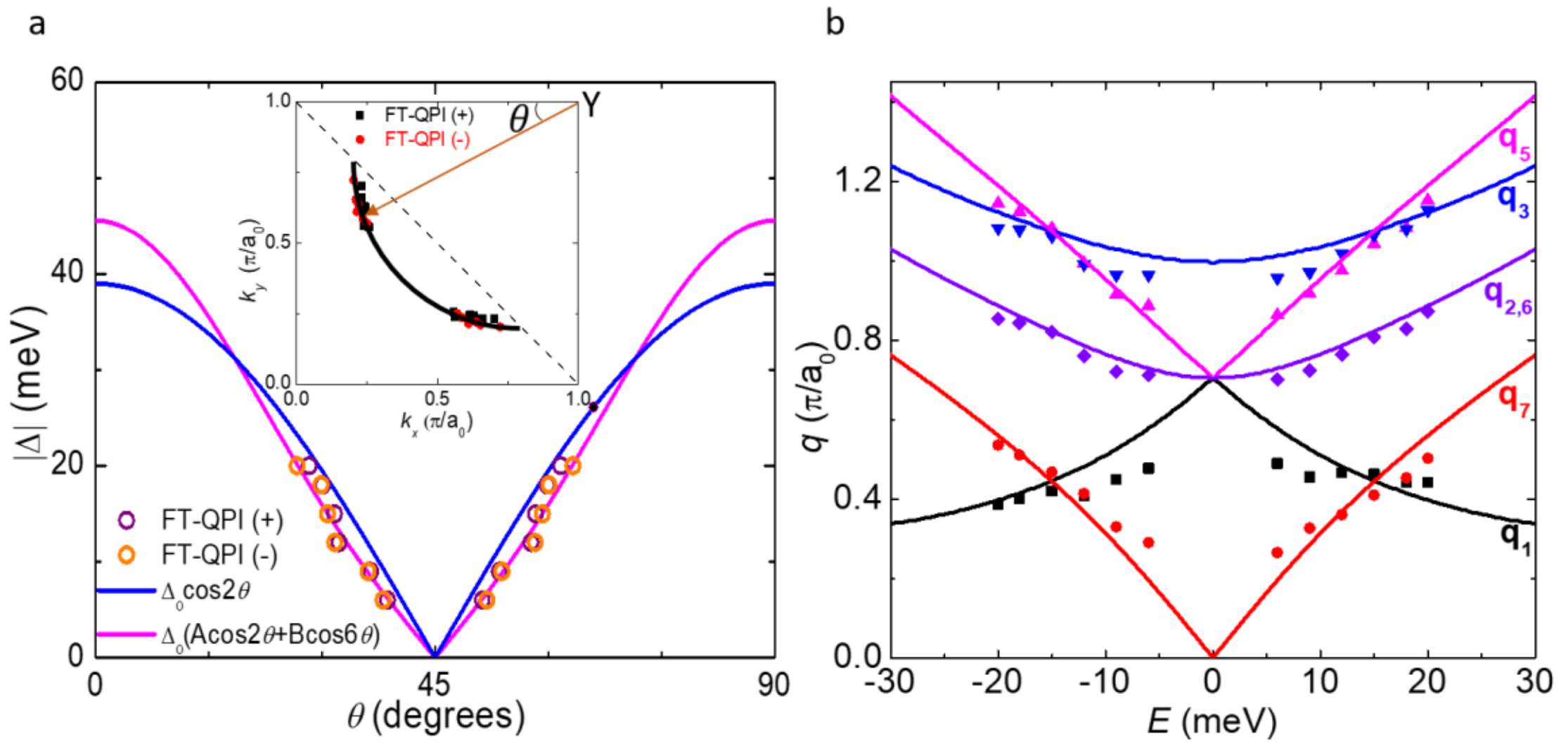

**Fig. 4** Energy dispersions for characteristic scattering wave vectors in different kinds of plots. **a** A plot of the superconducting gap $\Delta(\theta)$ and two fitting curves by different *d*-wave gap functions. The inset in **a** shows the positions of the octet terminals determined by $\mathbf{q_1}$ and $\mathbf{q_7}$ measured at various energies. The solid line shows the Fermi surface from the fitting to these positions by a circular arc, and it is cut off by the dashed line. The angle $\theta$ for each octet terminal is defined in the inset and about Y point ($\pi$, $\pi$). The gap value $\Delta$ for each scattering terminals is equivalent to the energy at which the QPI data are measured. **b** Energy dependent scattering wave vectors taken from the FT-QPI data (excluding $\mathbf{q_4}$ due to very weak intensity). The solid lines represent the theoretical predictions based on the Fermi surface and the *d*-wave gap function with high order obtained in **a**.

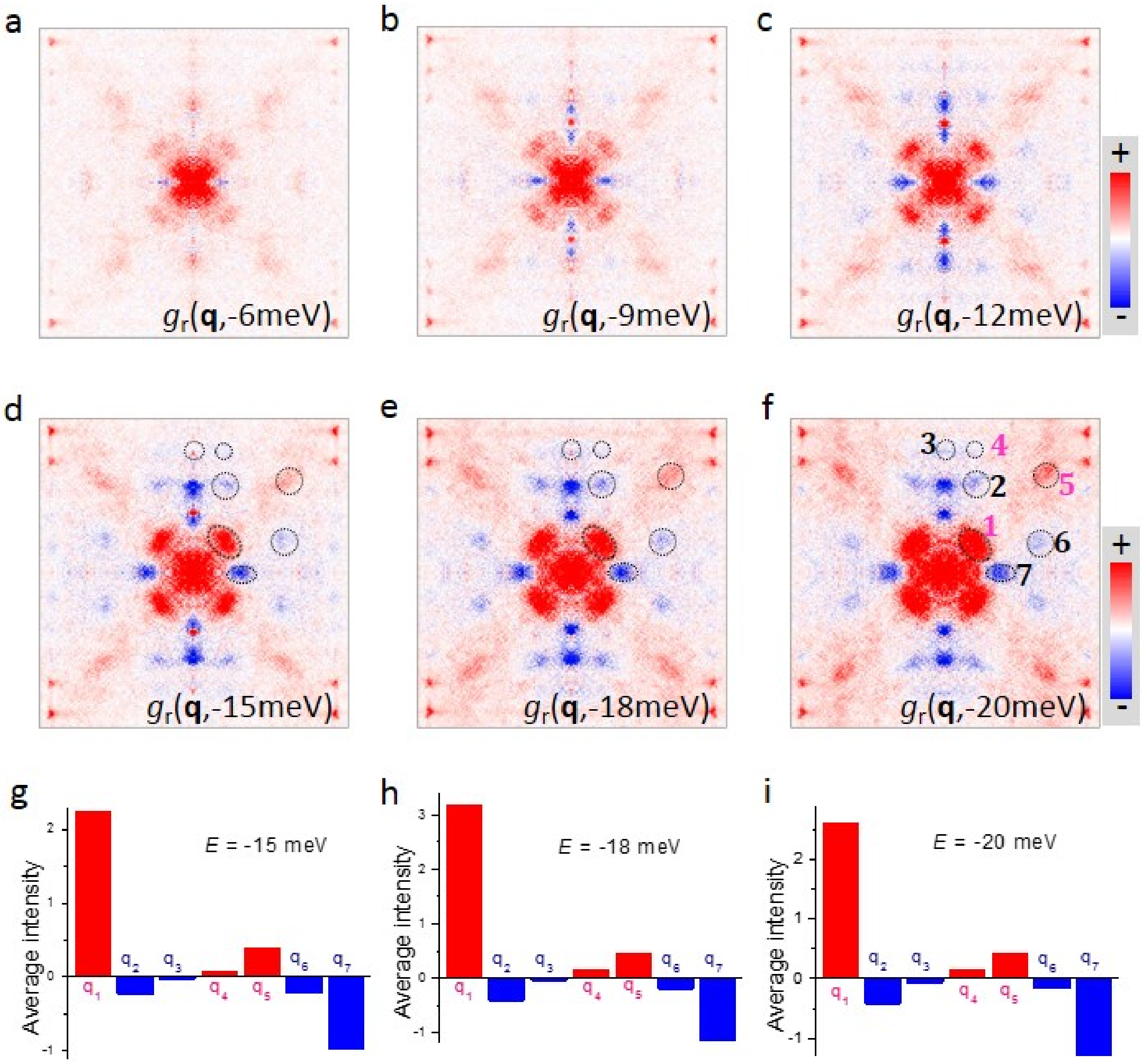

**Fig. 5** Experimental data by multi-IBS-QPI method. **a**-**f** PR-QPI images obtained directly from the FT-QPI patterns measured at different energies without multi-impurity phase correlation. Each number in **f** denotes the position of each characteristic scattering wave-vectors. **g**-**i** Integrated and averaged intensity per pixel of PR-QPI signal for each **q** at −15, −18, and −20 meV, respectively. The areas for the integral are marked by dashed circles or ellipses in **d**-**f**.

# Supplementary Information

# Directly visualizing the sign change of *d*-wave superconducting gap in $Bi_2Sr_2CaCu_2O_{8+\delta}$ by phase-referenced quasiparticle interference

Qiangqiang Gu, et al.

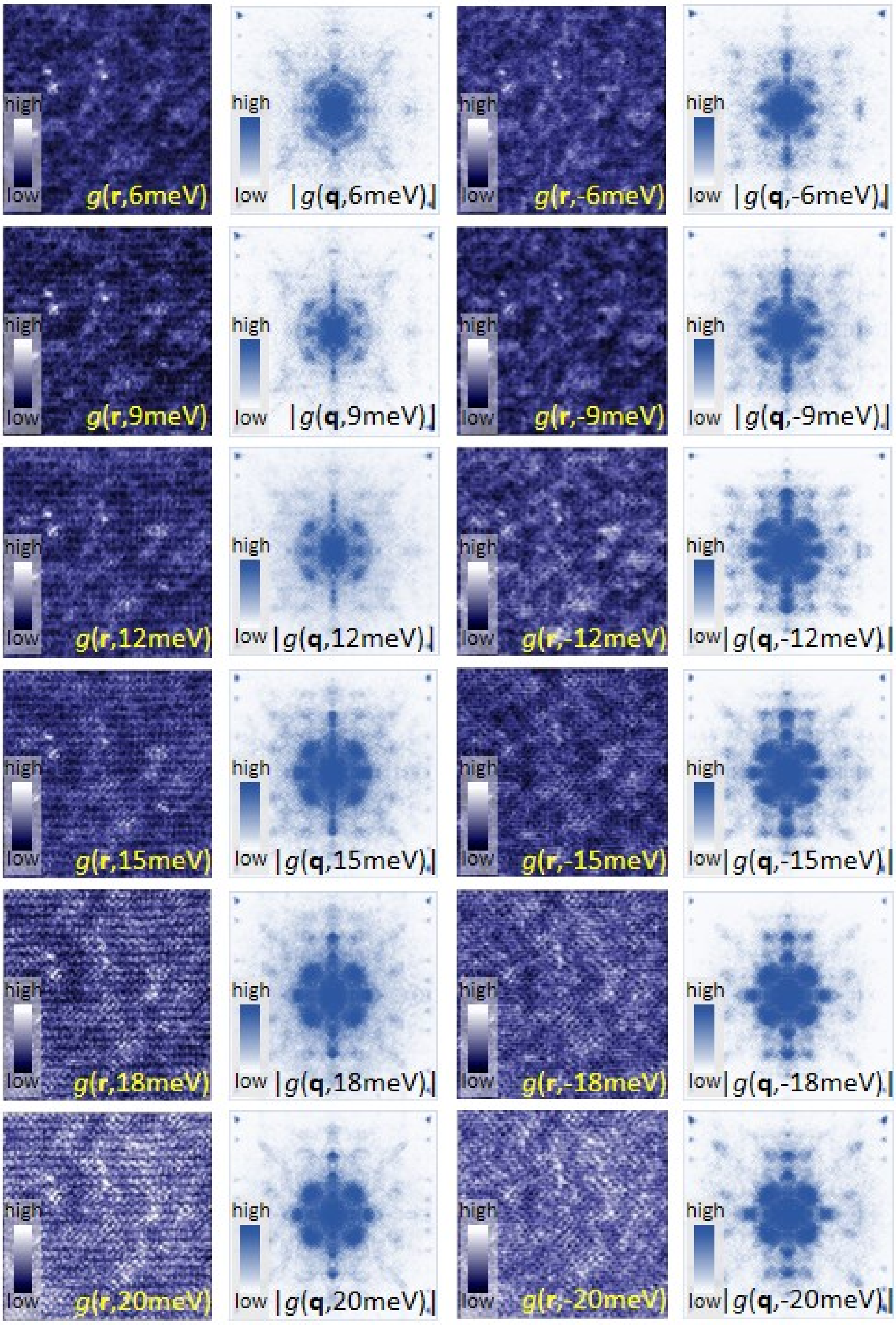

**Supplementary Fig. 1** Measured QPI images and corresponding FT-QPI patterns at different energies on Bi-2212.

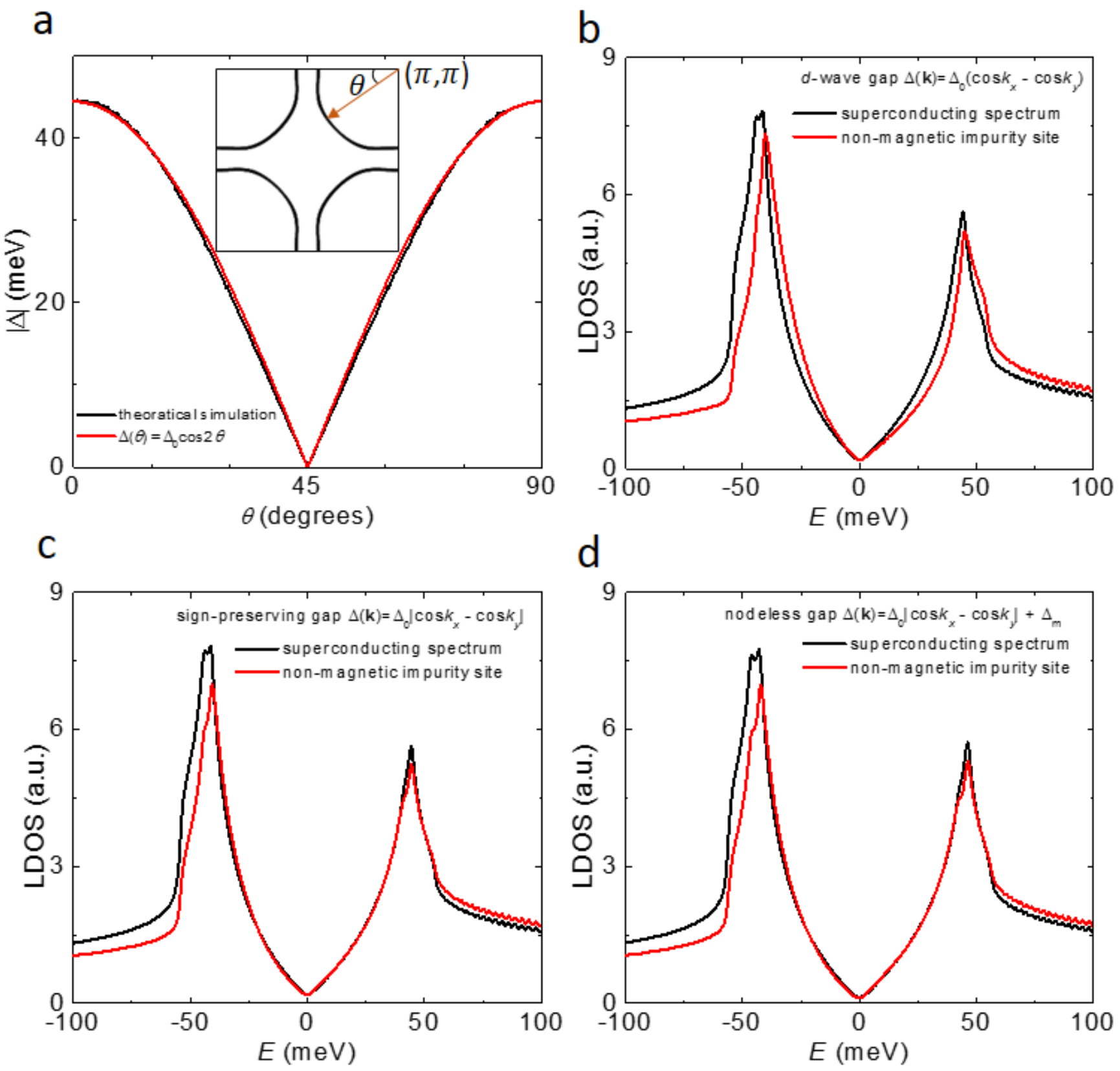

**Supplementary Fig. 2** Theoretical calculated tunneling spectra in different superconductors. **a** Fermi surface and gap function calculated for a *d*-wave superconductor. The inset in **a** shows the calculated Fermi surface and the definition of the parameter angle $\theta$. The *d*-wave gap function $\Delta(\mathbf{k}) = \Delta_0(\cos k_x - \cos k_y)$ with $\Delta_0$= 23 meV is assigned to the Fermi surface, and the resultant angle $\theta$ dependent superconducting gap along the Fermi surface is shown in **a**. The gap function along the Fermi surface is consistent with a *d*-wave function $\Delta(\theta) = 44.5\cos 2\theta$ meV. **b** The tunneling spectra in the place with or without non-magnetic impurity in a *d*-wave superconductor with gap function of $\Delta(\mathbf{k}) = \Delta_0(\cos k_x - \cos k_y)$. **c** The tunneling spectra in the place with or without non-magnetic impurity in a superconductor with nodal but sign-preserved gap function of $\Delta(\mathbf{k}) = 23|\cos k_x - \cos k_y|$ meV. **d** The tunneling spectra on the place with or without non-magnetic impurity in a superconductor with a nodeless gap function of $\Delta(\mathbf{k}) = (23|\cos k_x - \cos k_y| + 2)$ meV. The scattering scalar potentials are all set to $V_s = 20$ meV for the impurities in this figure.

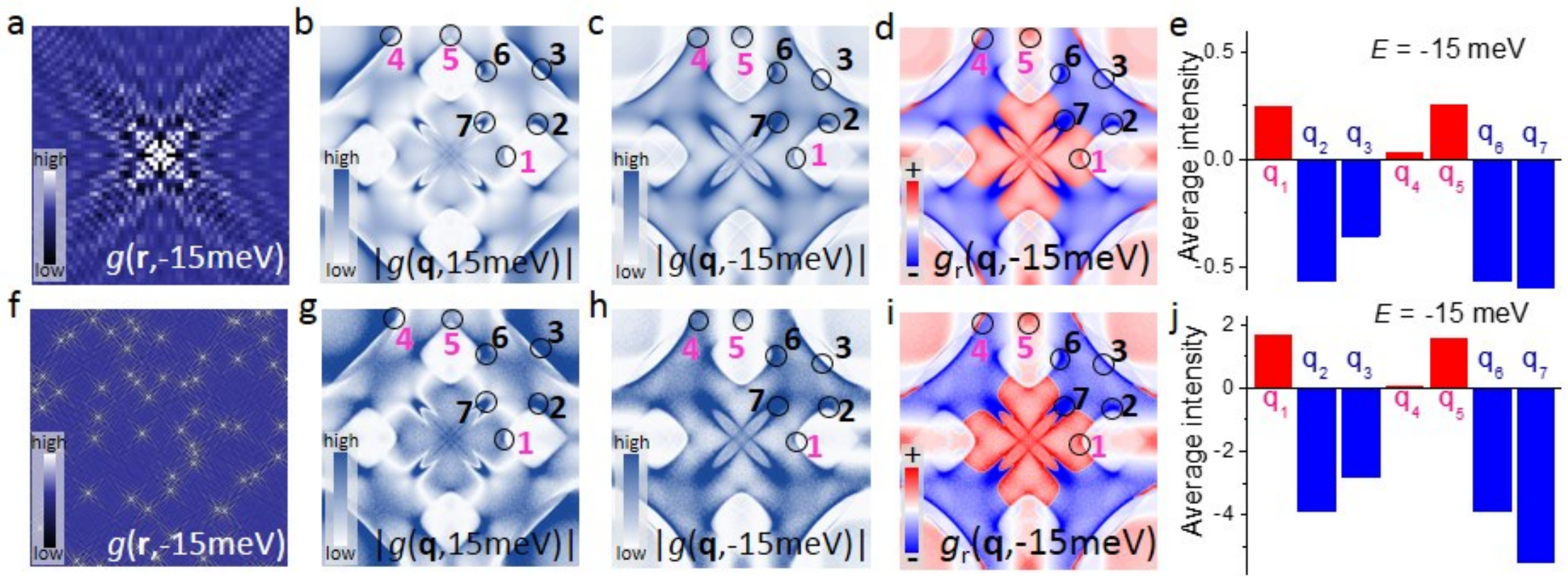

**Supplementary Fig. 3** Theoretical simulated PR-QPI results with a *d*-wave gap function. **a** Simulated LDOS around a single non-magnetic impurity at −15 meV with dimensions of 60×60 atom lattice. The *d*-wave gap function $\Delta(\mathbf{k}) = \Delta_0(\cos k_x - \cos k_y)$ with $\Delta_0 =$ 23 meV. The impurity is set to be at the center of the image. The related tunneling spectra are shown in Supplementary Fig. 2b. **b**,**c** FT-QPI patterns for the single impurity at $\pm$15 meV. **d** The resultant PR-QPI image $g_r$(**q**, −15 meV) calculated by IBS-QPI method from the simulation results in **b** and **c**. **e** Average intensity per pixel for the characteristic scattering spots in **d**. One can find that the $g_r$(**q**, −15 meV) values near gap-sign-preserved scattering vectors $\mathbf{q_1}$, $\mathbf{q_4}$, and $\mathbf{q_5}$ are positive, while the values near gap-sign-reversed scattering vectors $\mathbf{q_2}$, $\mathbf{q_3}$, $\mathbf{q_6}$, and $\mathbf{q_7}$ are negative. **f**-**j** The corresponding simulation results for the case of multiple impurities. The 60 impurities are the same as the one in **a**, and they are randomly distributed in an area with dimensions of 512×512 atom lattice as shown in **f**.

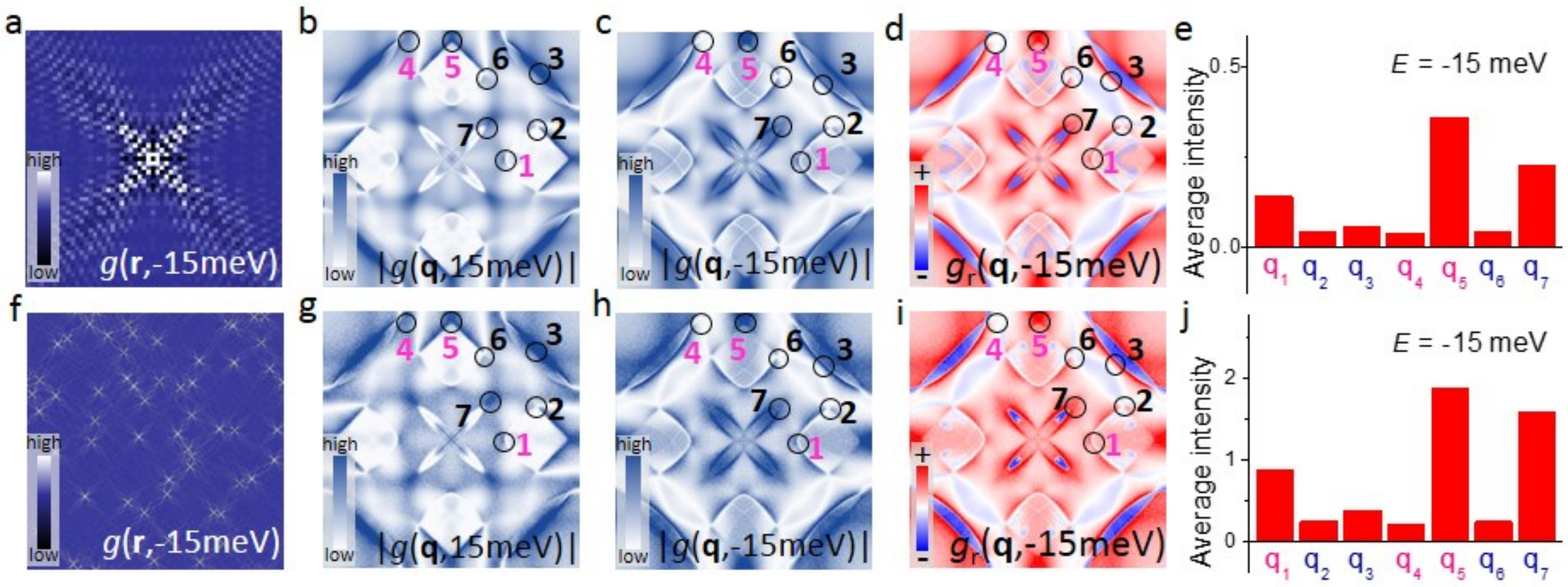

**Supplementary Fig. 4** Theoretical simulated PR-QPI results with a sign-preserved nodal gap function. **a** Simulated LDOS around a single non-magnetic impurity at −15 meV with dimensions of 60×60 atom lattice. The gap function used in the calculation is $\Delta(\mathbf{k}) = \Delta_0|(\cos k_x - \cos k_y)|$ with $\Delta_0$= 23 meV, which is a nodal but sign-preserved superconducting gap. The impurity is set to be at the center of the image. The related tunneling spectra are shown in Supplementary Fig. 2c. **b**,**c** FT-QPI patterns for the single impurity at $\pm$15 meV. **d** The resultant PR-QPI image $g_r$(**q**, −15 meV) calculated by IBS-QPI method from the simulation results in **b** and **c**. **e** Average intensity per pixel for the characteristic scattering spots in **d**. One can find that the $g_r$(**q**, −15 meV) values near all characteristic scattering vectors are positive. **f**-**j** The corresponding simulation results for the case of multiple impurities. The 60 impurities are the same as the one in **a**, and they are randomly distributed in an area with dimensions of 512×512 atom lattice as shown in **f**. The locations of impurities are the same as the ones in Supplementary Fig. 3f.

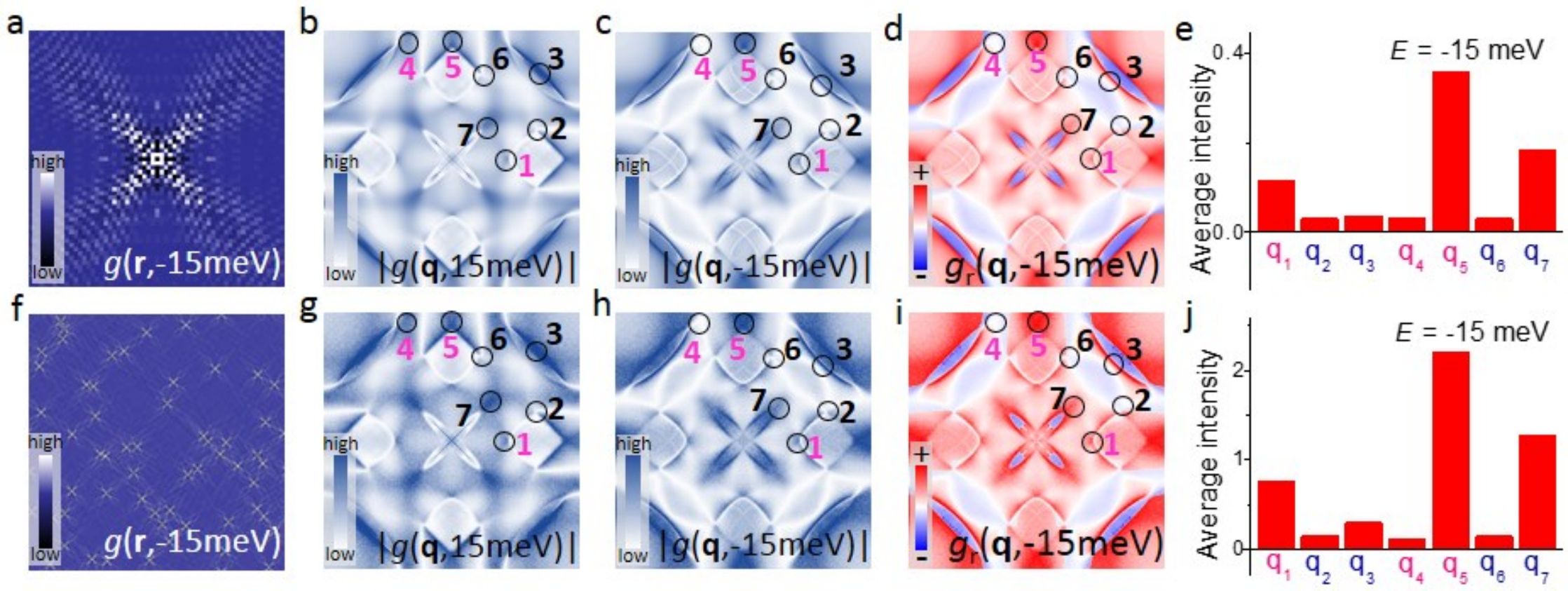

**Supplementary Fig. 5** Theoretical simulated PR-QPI results with a sign-preserved nodeless gap function. **a** Simulated LDOS around a single non-magnetic impurity at −15 meV with dimensions of 60×60 atom lattice. The gap function used in the calculation is $\Delta(\mathbf{k}) = \Delta_0|\cos k_x - \cos k_y| + \Delta_m$ with $\Delta_0$= 23 meV, $\Delta_m$= 2 meV, which is a sign-preserved nodeless superconducting gap. The impurity is set to be at the center of the image. The related tunneling spectra are shown in Supplementary Fig. 2d. **b**,**c** FT-QPI patterns for the single impurity at ±15 meV. **d** The resultant PR-QPI image $g_r(\mathbf{q}$, −15 meV) calculated by IBS-QPI method from the simulation results in **b** and **c**. **e** Average intensity per pixel for the characteristic scattering spots in **d**. One can find that the $g_r(\mathbf{q}$, −15 meV) values near all characteristic scattering vectors are positive. **f**-**j** The corresponding simulation results for the case of multiple impurities. The 60 impurities are the same as the one in **a**, and they are randomly distributed in an area with dimensions of 512×512 atom lattice as shown in **f**. The locations of impurities are the same as the ones in Supplementary Fig. 3f.